\documentstyle[12pt]{article}
\include{epsf}
\def\lo{\langle 0 |}
\def\ro{ | 0 \rangle }

\def\la{\langle }
\def\ra{ \rangle }

\newcommand{\beq}{\begin{equation}}
\newcommand{\eeq}{\end{equation}}
\newcommand{\bea}{\begin{eqnarray}}
\newcommand{\eea}{\end{eqnarray}}

\begin{document}
                                        \begin{titlepage}
\begin{flushright}
hep-ph/9711398
\end{flushright}
\vskip1.8cm
\begin{center}
{\LARGE
 Can $ \theta /N  $ Dependence for Gluodynamics \\       
\vskip0.7cm
be Compatible with $ 2 \pi $ Periodicity in $ \theta $ ?   \\
}         
\vskip1.5cm
 {\Large Igor~Halperin} 
and 
{\Large Ariel~Zhitnitsky}
\vskip0.5cm
        Physics and Astronomy Department \\
        University of British Columbia \\
        6224 Agricultural Road, Vancouver, BC V6T 1Z1, Canada \\ 
        {\small e-mail: 
higor@physics.ubc.ca \\
arz@physics.ubc.ca }
\vskip1.0cm
{\Large Abstract:\\}
\end{center}
\parbox[t]{\textwidth}{ 
In a number of field theoretical models the vacuum angle $ \theta $ 
enters physics in the combination $ \theta/N $, where $ N $ stands
generically for the number of colors or flavors, in an apparent 
contradiction with the expected $ 2 \pi $ periodicity in $ \theta $.
We argue that a resolution of this puzzle is related to the 
existence of a number of different $ \theta $ 
dependent sectors in a finite volume formulation, which can not be seen
in the naive thermodynamic limit  
$ V \rightarrow \infty $. 
It is shown that,
when the limit $ V \rightarrow \infty $ is properly defined, physics
is always $ 2 \pi $ periodic in $ \theta $ for any 
integer, and even rational, values of N, with 
vacuum doubling at certain values of $ \theta $.
We demonstrate this phenomenon in both the multi-flavor 
Schwinger model
with the bosonization technique, and four-dimensional gluodynamics
with the effective Lagrangian method. The proposed mechanism  
works for an arbitrary gauge group. 

}

\vspace{1.0cm}

                                                \end{titlepage}

\section{Introduction}

Very soon after the discovery of instantons \cite{BPST} in
Yang-Mills (YM) theory it has become clear \cite{Jac} that
the latter possesses a hidden parameter $ \theta $ whose effects may
show up due to a non-trivial topological structure of the theory.
It is believed that $ \theta $ is an angular parameter, i.e. physics
is periodic in $ \theta $ with period $ 2 \pi $. 
In particular, the values $\theta=0$ and $\theta=2\pi$ 
correspond to one 
and the same 
theory. This is a direct consequence of the topological classification
of the  gauge theories, which is based on the assumption that fields
are smooth and regular. 

The fact of the existence of this new fundamental constant has 
immediately posed two difficult questions related to 
the so-called $U(1)$ and strong CP
problems. It has been argued by `t Hooft \cite{tH} that instantons
may lead to a resolution of the U(1) problem \cite{Wein}. Later,
Witten and Veneziano \cite{Wit} have found,  
within the large $ N_c $ approach, that 
physics should depend on $ \theta $ 
through the combination $\theta/N_c$ in order for 
the $U(1)$ problem to be solved. On the 
other hand, a non-zero value of 
$ \theta $ implies \cite{SVZ}, \cite{Crewther} a violation  
of CP invariance in strong interactions, which is not observed 
experimentally.  
At present there is no convincing
theoretical argument as to why 
$ \theta $ is so small. Most likely, the strong CP problem 
can not be 
solved  within the strong interaction sector of the 
Standard Model, and will not be discussed here. 

In the present work we address a different, but  related, question.
As we mentioned earlier, a resolution of the $U(1)$ problem 
suggests that $ \theta $ dependence comes in 
the combination $\theta/N_c$.
To show this, we recall the famous Witten-Veneziano  
relation \cite{Wit}
\beq
\label{2}
f_{\eta'}^2 m_{\eta'}^2 =  12 \, \frac{ \partial^2 E_{vac} }{
\partial \theta^2 } \; \;   \; , \; \; \; E_{vac}\sim N_c^2 f(\theta)
\eeq
  Here $ f_{\eta'} $ is the $ \eta' $ residue 
and $E_{vac} $
is the energy of the YM vacuum, which is  proportional 
to $N_c^2$ in large $N_c$ limit.
 A resolution of the $U(1)$ problem suggests \cite{Wit} that 
$f_{\eta'}^2 m_{\eta'}^2\sim N_c\times 1/N_c = O(N_{c}^0) $.
Therefore, the right hand side of Eq.(\ref{2})  
should also be $  O(N_{c}^0)$.
This is exactly the case provided  we accept that the  
function $ f $ in formula (\ref{2})
is actually a function of the variable   $\theta/N_c$, rather 
than  of $\theta$ itself.
This completes the standard argument showing
that the  $\theta$ 
dependence should come in the combination $\theta/N_c$ only.
There exist many other arguments (based on analyses of $ 2 D $ 
$ CP^{N-1} $ model, SUSY theories, etc.) which support the 
conclusion that the $ \theta $ angle always enters physics
in a combination $ \theta / N $ where $ N $ is the number of 
``colors" or ``flavors", depending on the model considered. 

The question we want to raise (and attempt to answer)  can be 
formulated as follows.
How can one reproduce the $2\pi$ periodicity in $ \theta $ 
(which is a strict constraint following  from the 
topological
classification)
for {\bf all} physical quantities  if the  
  {\bf same}
 quantities  depend on $ \theta $ through the specific combination
 $\theta/N_c$ ? 
The answer to this question is well known in 
SUSY models \cite{SV},\cite{KS},
where it was shown that the $2\pi$ periodicity is recovered 
when a discrete number of vacuum 
states ($N$ for the $SU(N)$ group, $N-2$ for the $SO(N)$
group, etc.) is taken into account. 
The existence  of these states
in SUSY models is a consequence of spontaneous  
breaking of the discrete chiral symmetry $
Z_{2N_c} \rightarrow Z_2 $ (for the SU(N) gauge 
group), which shows up via
a formation of the gluino condensate. In this case, the 
different vacua
are labeled by the $\theta$ angle as well as 
a discrete parameter $k=0, 1,... N-1$
such that the chiral condensate depends
on these parameters as 
 $\la\bar{\lambda} \lambda \ra\sim
\exp{(\frac{i\theta +2\pi k}{N})}$.
Therefore,  when $\theta $ varies 
continuously from $0$ to $2\pi$, $ N_c $ distinct and disconnected
Bloch type vacua undergo a cyclic permutation: 
 the 
first state
becomes the second one, and so on. All physical quantities are 
periodic in $ \theta $ with periodicity $ 2 \pi $, as these 
vacua can be just relabeled by the substitution   
$k\rightarrow k-1$ after the shift $ \theta \rightarrow 
\theta + 2 \pi$,   
keeping physics intact.
We should note that such a picture is believed to be correct
for an arbitrary gauge group, irrespective of the 
existence of the center of the group\footnote{A related question
on  a role of the 
torons \cite{hooft},\cite{ARZ1},\cite{Shif},\cite{Smilga},
\cite{KSS},
which are field configurations  with a fractional topological
charge, is not addressed in this paper, see Ref.\cite{Smilga} for a 
list of related problems and discussions.}.

In non-supersymmetric models  such a scenario apparently
can not be realized because no discrete symmetry which could 
lead to such degenerate vacua exists 
in a pure gauge theory. The main goal of the present paper
is to argue that the pattern of the $ \theta $ 
dependence in usual, non-supersymmetric 
YM theory (gluodynamics) is to some extent reminiscent of 
what happens in SUSY models,  
 although there are important differences between these two cases.
 
What will be shown is 
that   
a discrete number of states, whose presence is 
 crucial for the aforementioned mechanism to work,  does 
exist 
in non-supersymmetric gluodynamics when we consider the theory
in a large, albeit finite, volume $ V $. These states represent  
 in this case local extrema of an effective potential, and 
have different energies. 
Under the shift $ \theta \rightarrow \theta + 2 \pi $, they 
transform to each other by a cyclic permutation, while some two 
of them cross in energy at certain values of $ \theta  
$ \footnote{ This picture of the $ \theta $ dependence
is similar to the one advocated by Crewther
\cite{Cr}, Witten \cite{Wit2}, and Di Vecchia 
and Veneziano \cite{VV} for QCD with $ N_f $ light flavors.}. 

Thus, the periodicity in $ \theta $ with period $ 2 \pi $ is 
restored in this finite volume theory. However, when the 
thermodynamic limit $ V \rightarrow \infty $ is performed for 
a generic value of $ \theta $, 
only one state of lowest energy can be seen, as all other states 
have higher energies and therefore drop out in the standard
definition
\beq
\label{e}
E_{vac}( \theta) = - \lim_{V \rightarrow \infty} \frac{1}{V}
\log Z \; \; , \; \;  \theta \; fixed \; 
\eeq
 (This is in drastic contrast with the SUSY case where
all $ N_c $ vacua have the same vanishing energy, and thus all
survive the $ V \rightarrow \infty $ limit.)
On the other hand, due to the superselection rule
different states do not communicate to 
each other (and are absolutely
stable), and therefore the fact of existence of additional
higher energy states could be safely neglected, in
agreement with Eq.(\ref{e}), for all physical
problems except for one. Namely, retaining all these states 
is necessary for the analysis of periodicity in $ \theta $. 
The values $ \theta $ and $ \theta + 2 \pi $ are physically
equivalent for this set of states as a whole. 
Thus, the fact that the $ \theta $ 
dependence comes in the combination $ \theta/ N_c $ in usual 
$ V = \infty $ formulae has nothing to do with the problem of 
periodicity in $ \theta $, as those formulae refer to one 
particular state out of this set.
As for any fixed $ \theta $ the information
on all additional states is lost in Eq.(\ref{e}), in what follows 
this phenomenon will be referred to  
as a ``non-commutativity" of 
the thermodynamic limit $  V \rightarrow \infty $  
and the shift $ \theta \rightarrow \theta + 2 \pi $.   
Bearing in mind that the states
of the set do not interact owing to the superselection rule, 
and to have a correspondence with the standard definition
(\ref{e}), in what follows we call the true vacuum state a 
state of lowest energy (for a fixed $ \theta $) among this set.
When defined in this way, the physical vacuum is periodic 
in $ \theta $ with period $ 2 \pi $, but for different intervals of
the values of $ \theta $ 
we are talking 
about a different state from the set as a true vacuum.
At certain values of $ \theta $ an exact two-fold degeneracy 
in this set results in vacuum doubling in the limit
$ V \rightarrow \infty $ (Dashen phenomenon
\cite{Dash}).
 
As a warm-up example, we first discuss in Sect.2 the multi-flavor 
Schwinger model. Using the bosonization approach, we show 
that the $ \theta $ dependence is realized in this model 
in the way just described. An important role of an integer valued 
Lagrange multiplier field, ensuring the quantization of the 
topological charge, is clarified.

The rest of the paper is devoted to the problem of $ \theta $ 
dependence in four dimensional YM theory. To this end, the 
knowledge of an infinite series of zero momentum 
correlation functions of the topological density $ G_{\mu \nu} 
\tilde{G}_{\mu \nu} $ 
is required. We discuss in Sect.3 methods for getting
 information
of this sort by matching short distance and 
large distance properties of the theory (a related 
discussion can be found in the Appendix). We then construct in 
Sect.4 an effective Lagrangian (more precisely, effective potential)
as the (Legendre transform of) generating functional for zero momentum
correlation functions of the marginal operators $  G_{\mu \nu} 
\tilde{G}_{\mu \nu} $ and $ G_{\mu \nu} 
G_{\mu \nu} $. We shown that an integer valued 
Lagrange multiplier field should be introduced in this effective 
potential to ensure a single-valuedness and boundness from below.
The presence of this field imposes global quantization 
conditions on the fields of the effective theory, which 
reflect the topological charge quantization in the original YM 
action.
Sect. 5 deals with the minimization of an ``improved" effective 
potential obtained by adding this Lagrangian multiplier 
field to the theory. 
We show that this procedure yields the above  
picture of $ 2 \pi $ 
periodicity in $ \theta $, in close analogy 
to what we find in the multi-flavor Schwinger 
model. Conclusions and some discussion
are presented in final Sect.6.

\section{ $ \theta $ dependence in $N$-flavor Schwinger model }

The main goal of this section is to illustrate most 
essential technical tools, needed to address the problem
of $ \theta $ dependence in YM theory, on a simple toy 
model. The multi-flavor Schwinger model
nicely serves this purpose. It exhibits the above pattern of 
$ \theta $ dependence and can be analysed in quite a
straightforward manner. Though most of the results presented
in this section are not new, we hope that this 
discussion may help to understand the mechanism 
ensuring the 
  $2\pi$ periodicity in $\theta$ in a more complicated case of 
four dimensional YM theory.
 
 As is known \cite{Coleman},
\cite{Hosotani}, in the $ N $ flavor Schwinger model 
   physics depends on $\theta$ through
the combination $\theta/N$. At the same time, all physical
quantities are periodic functions of $ \theta $ with period 
$ 2 \pi $. We here wish to discuss the way in which
these two apparently contradictory facts become 
compatible. In what follows 
 we reproduce some of the results of Refs.\cite{Coleman},
\cite{Hosotani} using an approach which emphasizes a key role
of an integer-valued Lagrange multiplier field and  
thermodynamic limit procedure in this problem.

We start with the standard Euclidean action of the bosonized 
one flavor Schwinger
model. It takes the form
\beq
\label{a1}
 S_E=\int d^2x[\frac{1}{2g^2}F^2+\frac{1}{2}
(\partial_{\mu}\phi)^2+
\frac{i}{\sqrt{\pi}}F\phi-\mu m_q\cos(\sqrt{4\pi}\phi)
+\frac{iF\theta}{ 2\pi}],
\eeq
where $F=\frac{1}{2}\epsilon_{\mu\nu}F_{\mu\nu}$ and $m_q$ 
is a quark mass.
In obtaining this formula we used the standard boson-fermion 
correspondence
\beq
\label{a2}
\bar{\psi}\gamma_{\mu}\psi \rightarrow\frac{1}{\sqrt{\pi}}
\epsilon_{\mu\nu}\partial_{\nu}\phi \; ,
~~~~~~~~~\bar{\psi}\psi\rightarrow -\mu\cos(\sqrt{4\pi}\phi).
\eeq
The key observation is the following. 
We would like to impose explicitly
a global constraint ensuring that
the topological charge, which is determined by the integral
$\int d^2x\frac{F}{ 2\pi}$, can take only integer values.
This constraint can be imposed by introducing
an integer-valued Lagrange
multiplier variable $n$ such that 
the partition function of the theory is {\bf defined } as the 
sum over $n$:
\beq
\label{a3}
Z= \sum_n\int DFD\phi e^{-S_E+in2\pi\int d^2x\frac{F}{ 2\pi}}.
\eeq
It is clear that we have   done nothing wrong by defining
the partition function in this way, because we introduced only 
a phase
multiplier which always equals $1$ for integer topological 
charges. This 
procedure brings
a divergent normalization factor $\sum 1$ which is irrelevant anyhow.

Two remarks are in order.
First, the constraint $\int d^2x\frac{F}{ 2\pi}=l,~ l=0,\pm 1, 
\pm 2,...$
is automatically satisfied due to the identity:
\beq
\label{a4}
 \sum_n \exp \left( i2\pi n\int d^2x\frac{F}{ 2\pi}
 \right) =\sum_l \delta \left( \int d^2x\frac{F}{2\pi}-l \right) \; .
\eeq
As for the second, and  most important, remark: the 
$2\pi$ periodicity 
in $\theta$
is explicitly seen from the general expression (\ref{a3}). Indeed, 
a  shift
in $\theta$ by $2\pi$ in Eq.(\ref{a1}) can be compensated for by a 
shift in
$n:~~n\rightarrow n+1$ such that the partition function 
(\ref{a3}) is unchanged.
 The crucial point 
to make this 
mechanism work 
is the definition (\ref{a3}) of the partition function $Z$
with the prescription 
of   summing  over all $n$. Such a definition was  suggested earlier 
(with quite a different motivation)
for the Schwinger
model by Smilga \cite{Smilga1}, and for SUSY models
(with purposes similar to ours)  by Kovner and Shifman \cite{KS}.

To study the vacuum structure of the theory,  
we expand, following Ref.\cite{Smilga1},
the fields $F(x)$ and $\phi(x)$ in the series over spherical 
harmonics (a compactification on a manifold of volume $ V $ 
is implied)
$F(x)=\sum F_{lm}Y_{lm}(\Omega),~~\phi(x)=\sum 
\phi_{lm}Y_{lm}(\Omega)$ and 
  keep only the zero $F_0$  mode in what follows. 
This harmonic $F_0=\frac{2\pi l}{V} $
is fixed by the constraint (\ref{a4}), with $V$ 
being the total volume of the system. 
Integrating over $DF_0$  and discarding non-zero harmonics
of the $ \phi $ field, which are irrelevant for the
present discussion, we obtain for the partition function
\beq
\label{a5}
Z\sim \int_{\frac{-\sqrt{\pi}}{2}}^{\frac{+\sqrt{\pi}}{2}}
 D\phi_0\sum_{l=-\infty}^{+\infty}
 \exp \left\{ -\frac{2\pi^2l^2}{Vg^2}+il( \sqrt{4\pi}\phi_0+\theta)
+ V m_q \mu \cos ( \sqrt{4 \pi} \phi_0 )  \right\} \; .
\eeq
The $2\pi$ periodicity in $\theta$
is explicitly seen in this representation. However, in order
to discuss the thermodynamic limit, it is more convenient 
to use an alternative (dual) representation for the same expression 
(\ref{a5}):
\beq
\label{a6}
Z\sim \int_{\frac{-\sqrt{\pi}}{2}}^{\frac{+\sqrt{\pi}}{2}}
 D\phi_0\sum_{k=-\infty}^{+\infty}
 \exp \left\{ -\frac{g^2V}{2\pi}(  \phi_0-k\sqrt{\pi}+
\frac{\theta}{\sqrt{4\pi}})^2 
+ V m_q \mu \cos ( \sqrt{4 \pi} \phi_0 ) \right\} \; ,
\eeq
where we have used the property of  
$\theta_3(\nu, x)$ 
function:
\beq
\label{a7}
 \theta_3(\nu, x)=\frac{1}{\sqrt{\pi x}}\sum_{k=-\infty}^{+\infty}
e^{-\frac{(\nu+k)^2}{x}}=
  \sum_{l=-\infty}^{+\infty}
 e^{-l^2\pi^2x +2il\nu\pi }.
\eeq
A generalization of this formula for the case of
 $N_f$   flavors  
with equal (and very small, $m_q \ll g$) masses 
can be achieved by replacing $\phi_0\rightarrow N_f\phi_0$
in Eq.(\ref{a6}) where, again, we keep only the relevant
   for the vacuum structure  part of the partition function:
 \beq
\label{a8}
Z \sim \int_{\frac{-\sqrt{\pi}}{2}}^{\frac{+
\sqrt{\pi}}{2 }}
 D\phi_0\sum_{k=-\infty}^{+\infty}
 \exp \left\{ -\frac{g^2V}{2\pi}(  N_f\phi_0-k\sqrt{\pi}+\frac{\theta}{
\sqrt{4\pi}})^2 
+ V m_q \mu N_f \cos ( \sqrt{4 \pi} \phi_0 ) \right\} 
\eeq
Formula (\ref{a8}) was derived earlier   \cite{Smilga1} in the limit 
$\theta=0, N_f=1$. 

Now we are ready to discuss the periodic properties of the 
partition function
(\ref{a8}) in the thermodynamic limit $g^2V\rightarrow\infty$
for the strong coupling regime $ m_q \ll g $. 
It is clear beforehand, without any calculations, that
$Z $ is a periodic function of $ \theta $ with 
period $2\pi$
for an arbitrary $N_f$,
due to summation 
over the integers $ k $ in 
Eq.(\ref{a8}).
 Now we wish to see explicitly how 
this periodicity works in Eq. (\ref{a8}) in the limit  
$g^2V\rightarrow\infty$.
For sufficiently small $\theta$, 
only one term  $k=0$  in the sum 
(\ref{a8}) survives
the thermodynamic limit
, such that
 \beq
\label{a9}
\sqrt{4\pi}\phi_0=-\frac{\theta}{N_f} \; \; , \; \; g^2V\rightarrow
\infty \; \; , \; \;  k=0 \; \; , \; \; 0 \leq \theta < \pi \; .
\eeq
The vacuum energy and topological density 
condensate for solution (\ref{a9}) are
\bea
\label{a10}
E_{vac}(\theta) & \sim &
- \cos(\sqrt{4\pi}\phi_0)
\sim -\cos \frac{\theta}{N_f} \nonumber \\  
\la \frac{ i F }{ 2 \pi} \ra & \sim &  \sin \frac{ \theta}{N_f}  \; \; , 
\; \; 0 \leq \theta < \pi \; .
\eea
When $\theta > \pi$, the next term with $k=1$ 
corresponds to a lowest energy state, such that
 \beq
\label{a11}
\sqrt{4\pi}\phi_0=\frac{2\pi}{N_f}
-\frac{\theta}{N_f} \; \; , \; \; g^2V\rightarrow\infty \; \; , 
\; \; k=1 \; \; ,  \; \; \pi < \theta  \leq 2 \pi  \; .
\eeq
The vacuum energy  and topological density 
condensate for solution (\ref{a11}) are
\bea
\label{a12}
E_{vac}(\theta) & \sim & 
- \cos(\sqrt{4\pi}\phi_0)
\sim-\cos(\frac{2\pi}{N_f}
-\frac{\theta}{N_f}) \nonumber \\
\la \frac{ i F }{ 2 \pi} \ra & \sim & - \sin ( 
\frac{2 \pi}{ N_f} -  \frac{ \theta}{N_f} ) \; \; , \; \; 
\pi < \theta \leq 2 \pi \; .
\eea
This pattern continues for arbitrary values of $\theta$.  
For the special case $\theta=\pi$, a cusp singularity develops, and 
we stay with two degenerate vacua which both 
survive the thermodynamic limit (for $ N_f = 2 $, this phenomenon  
was discussed long ago by Coleman
\cite{Coleman}). In this case the two vacua are
distinguished by the 
sign of the topological density condensate $ \la iF/(2 \pi) \ra $.
 
To summarize, we explicitly see that physics depends on $\theta$ 
for sufficiently small $\theta$  through the 
combination $\theta/N_f$. At the same time, the period of 
the $ \theta $ dependence is standard $ 2 \pi $.
This result is in agreement with a very different 
approach of Ref.\cite{Hosotani}.

We thus see that the following picture emerges :\\
1. The $2\pi$ periodicity in $ \theta $ holds for an 
arbitrary value of $N_f$,
even if $N_f$ is a rational, and not an integer,
 number (thought it is not a physical situation for 
the Schwinger model, 
 a similar formula for  four-dimensional
gluodynamics exhibits such a possibility).  The $2\pi$ periodicity 
in $ \theta $ is always restored due to the summation over 
integer-valued Lagrange
multiplier variable $n$ in the partition function (\ref{a3}).
      
2. If we take the thermodynamic limit for 
sufficiently small $\theta$ in Eq.(\ref{a8})
from the very beginning, we obtain
$\sqrt{4\pi}\phi_0=-\frac{\theta}{N_f}$ once and forever for 
an arbitrary value of
$\theta$. 
Proceeding this way, 
 we would not see other terms with $k=1, 2...$ in
Eq.(\ref{a8}), which are responsible for the restoration of $2\pi$ 
periodicity in $ \theta $, 
simply because they contribute zero to the partition
function in the thermodynamic limit.
This is exactly what happens when one starts
with the continuum formulation of the theory from the very beginning.
In this case all terms with $k=1, 2...$ are automatically discarded.
Therefore, the $ 2 \pi $ periodicity in $ \theta $, as a 
property of the whole set of solutions, can not be seen
in this formulation. The thermodynamic limit prescription
(\ref{e}) and the shift $ \theta \rightarrow \theta + 2 \pi $
are thus ``non-commutative".
 
3. For each given $\theta\neq \pi$ (
and  $m_q\neq 0$) there is one and only 
one physical vacuum in the 
thermodynamic limit. By convention (\ref{e}) we define this vacuum
as a state of lowest energy among the above set.
For $\theta= \pi$ there are  exactly
two degenerate states
which both contribute the partition function in the limit
$ V \rightarrow \infty $. In this sense, at the crossing point 
$ \theta = \pi $ the physics is non-analytic in 
$ \theta $. This non-analyticity can not be seen
in usual $ V = \infty $ formulae valid for small
$ \theta < \pi $, where a contribution of a lowest 
energy state only is retained. 
An order parameter 
which labels two degenerate states at $ \theta = \pi $
is 
the sign of the vacuum expectation value of operator
$\la \frac{iF}{2\pi}\ra\sim \pm 
m_q\sin(\frac{\pi}{N_f})$. 

4. As we mentioned above, the choice of the physical
vacuum as a lowest energy state is a matter of convention.
Owing to the superselection rule, any state in the discrete series
(e.g. 
a next-to-lowest state)
could serve as a vacuum as well. However, due to the permutational
symmetry of the whole set of states under the shift 
$ \theta \rightarrow \theta + 2 \pi $, any redefinition of this sort
will result in $ 2 \pi $ periodicity in $ \theta $ for such a 
vacuum in the physical limit $ V \rightarrow \infty $. 

In the next sections we will see that the very same picture of 
the physical $ \theta $ dependence seems to appear in four 
dimensional YM theory.

\section{Low energy theorems in gluodynamics}

In this section we start to describe steps which have to be 
done in YM theory to obtain an expression analogous to 
Eq.(\ref{a8}) for the Schwinger model (the vacuum contribution to
the partition function or, in other words, the vacuum energy).
This aim requires knowing the zero momentum part of the 
YM partition function, which can not be described within 
perturbation theory. It turns out that this object can be 
studied using some matching conditions ensuring consistency
of the large distance properties of the theory with its small 
distance behavior fixed by renormalizability and asymptotic 
freedom. These matching conditions are provided by anomalously 
broken symmetries through a set of
Ward identities\footnote{
In SUSY models, a similar use 
of anomalous Ward identities leads to
the well known Veneziano-Yankielowicz
effective Lagrangian \cite{VY}.} for 
zero momentum correlation functions of 
operators describing corresponding anomalies. This constitutes
what is known as low energy theorems in gluodynamics.
 Our aim in this section is to 
discuss the low energy theorems 
in YM theory in order to prepare a 
necessary input for the construction of an effective Lagrangian, which
will be carried out in Sect.4.
  
In what follows we need two Ward identities for zero momentum
correlation functions of spin 0 gluon currents in gluodynamics. 
For the scalar channel case, it was shown long ago  
by Novikov et.al. \cite{NSVZ} (NSVZ) that these correlation functions 
are fixed by renormalizability and conformal
anomaly in YM theory. Indeed renormalizability and 
dimensional transmutation ensure that any renormalized zero 
momentum correlation function of the $ d = D = 4 $ operator 
$ G^2  $ can only be of the form $ C \la 
\beta(\alpha_s)/(4 \alpha_s) G^2 
\ra $, where $ C $ is a numerical constant which depends on the 
correlation function considered, and  the renormalized 
vacuum expectation value $
\la \beta(\alpha_s)/(4 \alpha_s) G^2
\ra  \sim \Lambda_{YM}^{4} $ is the only mass scale in the theory, 
fixed by the conformal anomaly\footnote{The fact that terms explicitly
containing $ \Lambda_{YM}^{4} $ do not appear in correlation functions
of the operator $ G^2 $ was checked 
in Ref.\cite{NSVZ} using canonical methods with Pauli-Villars
regularization.}.
 For any 
given zero momentum correlation function of the field $ G^2 $, a 
value 
of the particular coefficient $ C $ can be found using the 
dimensional transmutation formula for renormalized vacuum expectation
value of operator $ O $ of canonical dimension $ d $, written in terms
of the bare coupling constant $ g_0 $ normalized at the 
cut-off scale $ M_R $:
\beq
\label{4}
\la O \ra = const \; \left[ M_R \exp \left(- \frac{8 \pi^2}{b
g_0^2} \right) \right]^d  \; ,
\eeq
where the one-loop $ \beta $-function, $ \beta(\alpha_s) = 
- b \alpha_{s}^2 /(2 \pi) $ 
 with
 $ b = (11/3) N_c $ and $ N_c $  stands for the number of colors, 
has been used.
 The NSVZ theorem \cite{NSVZ} (with the one-loop  
$ \beta $-function)
then follows by the 
differentiation of Eq.(\ref{4}), taken
for $ O =   - b \alpha_s / (8 \pi) G^2 $,
in respect to $ 1/ g_{0}^2  $. When the full $ \beta $-function 
is retained, it reads \cite{NSVZ}
\beq
\label{3}
\lim_{q \rightarrow 0 } \; i 
\int dx \, e^{iqx} \lo T 
\{ \frac{\beta(\alpha_s)}{4 \alpha_s} G^2 (x)  \,  
 \frac{\beta(\alpha_s)}{4 \alpha_s} G^2 (0) \} \ro  = 
-4  \la \frac{
\beta(\alpha_s)}{ 4 \alpha_s} G^2 \ra \; ,
\eeq
Note that the presence of the full $\beta$-function
in Eq.(\ref{3}) ensures the renormalization group invariance of 
both
sides of this relation. An infinitesimally small momentum transfer 
$ q_{\mu} $ is introduced in order to select a connected 
contribution
to the correlation function (\ref{3}). 
Eq.(\ref{3}) stands for the renormalized 
correlation function where ultra-violet 
divergent 
contributions are implied 
to be regularized and subtracted in both sides of 
of Eq.(\ref{3}), see the Appendix for a discussion on this point.
  
Arbitrary n-point functions of the trace of the energy-momentum
tensor 
\beq
 \la \sigma \ra = \la \frac{\beta(\alpha_s)}{4 \alpha_s} \,  G^2
\ra = \la \frac{- b \alpha_s }{8 \pi} \, G^2 \ra  + O(\alpha_{s}^2 ) 
\eeq 
can be obtained by further 
differentiating relation (\ref{4}) :
\beq
\label{5a}
i^n \int dx_1 \ldots d x_n \lo T \{ \sigma (x_1) \ldots \sigma
(x_n) \, \sigma (0) \} \ro = ( - 4)^n  \, \la \sigma \ra \; , 
\eeq
where, as in Eq.(\ref{3}), a limiting procedure 
of the vanishing momentum transfer $ q_{\mu} $ is implied.
Note that a regularization scheme in Eqs.(\ref{3}),(\ref{5a})
is assumed to be the same. 
 
 Let us now address zero momentum 
correlation functions of the topological density operator
in gluodynamics. As discussed in detail in the Appendix,
 the renormalized  
two-point function can be written as 
\beq
\label{5}
\lim_{ q \rightarrow 0} \; 
i \int dx \, e^{iqx} \lo T \left\{ \frac{\alpha_s}{8 \pi} 
G \tilde{G} (x)  \, 
\frac{\alpha_s}{8 \pi} G \tilde{G} (0) \right\}  \ro =    
 \xi^2  \, \la \frac{
\beta(\alpha_s)}{ 4 \alpha_s} G^2 \ra \; ,
\eeq
where $ \xi $ stands for a generally unknown numerical coefficient
(note that its $N_c $ dependence is expected to be 
$ \xi \sim N_{c}^{-1} $, in order to match Witten-Veneziano
\cite{Wit} resolution of the U(1) problem). 
When $ \theta = 0 $, Eq.(\ref{5}) is the 
only possible form compatible with both the conformal anomaly 
and Witten-Veneziano construction. We note  that the correlation 
function (\ref{5}) is defined via the path integral, i.e. with 
Wick type of the T-product. This definition ensures that 
the nonperturbative gluon condensate in Eqs.(\ref{3}),(\ref{5})
is the same quantity. Perturbative contributions are absent in 
Eq.(\ref{5}), see the Appendix.

In writing Eq.(\ref{5}) we have assumed that
it has the same form, i.e. covariant, for any 
small value of $ \theta $ (in the Appendix it is proved only 
for $ \theta = 0 $).  
Provided this is the case, the coefficient $ \xi^2 $ 
reiterates, analogously to 
Eq.(\ref{5a}), for all n-point correlation functions of $ G \tilde{G}
$, as can be seen by the formal differentiation 
of Eq.(\ref{5}) in respect to $ \theta $. There are three 
arguments 
in favor of correctness of this assumption. 
First, this requirement 
agrees with  the large 
$N_c $ line of reasoning due to Veneziano \cite{Wit} where one
finds 
that a coefficient standing in the two-point function of the 
topological density operator does reiterate in multi-point 
correlation functions. Second, this postulated covariance of 
Eq.(\ref{5}) in respect to $ \theta $ 
 goes through
a self-consistency check by agreement between two different 
calculations
of the $ \theta $ dependence of the vacuum energy, when one 
of them is obtained by a straightforward 
use of Eq.(\ref{5}) 
(see Eq.(\ref{7}) below), and another one is 
obtained directly from an effective potential (see Sect.5).
Third, such covariance of Eq.(\ref{5}) follows automatically
with an approach used in \cite{6}.

As for the numerical coefficient $ \xi $ in Eq.(\ref{5}), there
exist a few proposals to fix its value. One of them \cite{NSVZ}
is based on the hypothesis of the dominance of self-dual fields
in the YM vacuum, which suggests $ \xi = 2/b $. A different choice,
$ \xi = 4/(3 b) $, was advocated in our work \cite{6}
 using  
a one-loop connection between the conformal 
and axial anomalies in the theory with an auxiliary heavy fermion.
This line of reasoning was an extension of arguments used by 
K\"{u}hn and Zakharov \cite{KZ} to evaluate the proton matrix 
element $ \la p | G \tilde{G} | p \ra $. Some further discussion 
on these matters will be given in Sect.5. As different arguments
disagree on what the exact value of $ \xi $ is, 
in this paper we prefer to proceed with an unspecified parameter  
$ \xi $. Fortunately, it turns out that the fact of $ 2 \pi $ 
periodicity in $ \theta $ can be established without knowing 
a precise value of $ \xi $, with the only mild and reasonable 
assumption that the parameter $ \xi $ is a rational number.
In addition, keeping an unspecified value of $ \xi $ in Eq.(\ref{5})
makes it possible to study the $ \theta $ dependence for gauge 
groups other than SU(N), at least in principle (see Sect.5). 

As has been shown in \cite{6}, a combined use of 
relations (\ref{3}),(\ref{5}) enables us to calculate
the $ \theta $ dependence of the vacuum energy and topological
density condensate for any number of colors $ N_c $ and small 
values of the vacuum angle $ \theta $ by formal 
resummations of Taylor expansions in $ \theta $ for these 
objects. The resulting expressions read 
\beq
\label{7}
E_{v}(\theta) = \la \theta | - \frac{b \alpha_s}{32 \pi}  G^2
| \theta \ra =    \lo - \frac{b \alpha_s}{32 \pi}  G^2 \ro
\, \cos (2 \xi \theta ) \equiv
E_v  \cos (2 \xi \theta ) \; .
\eeq
for the 
vacuum energy (here $ E_v $ stands for 
the vacuum energy for $ \theta = 0 $ )
and 
\beq
\label{8}
\la \theta | \frac{ \alpha_s}{ 8 \pi} G \tilde{G} 
 | \theta \ra =  2 \xi E_v 
 \sin ( 2 \xi \theta ) \; .
\eeq
for the  topological density condensate (in \cite{6} the 
particular value $ \xi = 4/(3b) $ was used). These formulas
seem puzzling as they suggest a ``wrong" 
periodicity in $ \theta $ (remember that $ \xi \sim N_{c}^{-1} $) 
without any hint at possible
singular points $ \theta \sim \pi $, which could prevent us 
from making the shift $ \theta \rightarrow \theta + 2 \pi $.
This might force one to conclude that equations (\ref{7}),(\ref{8})
can not be correct on general grounds, even for small values of 
$ \theta $, and the whole derivation, leading to relations
 (\ref{7}),(\ref{8}), was in error.
However, as we just saw in the analysis of the Schwinger model, a 
fractional $ \theta $ dependence, implied in Eqs.(\ref{7}),(\ref{8}),
can be in perfect agreement with the $ 2 \pi $ periodicity in 
$ \theta $,
see Eq.(\ref{a10}).
 What will be argued below is 
that Eqs.(\ref{7}),(\ref{8}) do {\bf not} contradict the 
expected picture of $ 2 \pi $ periodicity of
physics in $ \theta $ with 
a singular level crossing point 
at $ \theta \sim \pi $, for any rational number $ \xi $ (including, of 
course, both aforementioned choices
$ \xi = 2/b $ \cite{NSVZ} or $ \xi = 4/(3 b) $ \cite{6}).
As we have found in the study of 
the Schwinger model, the key to understanding 
the $ \theta $ periodicity problem is the analysis
of a whole set of disconnected vacuum states.
It will be shown in Sect.5 that an accurate transition to 
the limit $   V \rightarrow \infty $ while keeping
all these states restores 
the correct periodicity and analyticity
structure of the $ \theta $ dependence, irrespective of 
a particular value $ \xi = any \; rational \; number $. 
 
\section{ Effective Lagrangian for gluodynamics}

The purpose of this section is to construct a 
low energy effective Lagrangian for gluodynamics, which
would contain all information provided by the low energy 
theorems in the scalar (\ref{3}) and pseudoscalar (\ref{5}) channels
including all multi-point correlation functions of operators 
$ G^2 $ and $ G \tilde{G} $, which can be obtained by differentiating
the two-point functions (\ref{3}) and (\ref{5}),
see e.g. Eq.(\ref{5a}).

Before proceeding with the presentation, we would like to pause
for a comment on the meaning of this effective Lagrangian. As 
there exist no Goldstone bosons in pure YM theory, no Wilsonian
effective Lagrangian, which would correspond to integrating out 
heavy modes, can be constructed for gluodynamics. Instead, one 
speaks in this case of an effective Lagrangian as a generating 
functional for vertex functions of the composite fields $ G^2 $ and 
$ G \tilde{G} $. Moreover, only the potential part of this 
Lagrangian can be found as it corresponds to zero momentum n-point
functions of $ G^2 , G \tilde{G} $, fixed by the low energy theorems.
(This effective potential still contains an ambiguity which will play 
an important role in what follows.) The kinetic part is not fixed 
in this way. Thus, such an effective Lagrangian is not very useful
for calculating the S-matrix, but is perfectly suitable for addressing 
the vacuum properties\footnote{  
Effective Lagrangians of this kind have been used 
in supersymmetric theories (see e.g. review 
papers \cite{SUSY}). In particular, the so-called 
Veneziano-Yankielowicz effective Lagrangian \cite{VY} has the 
meaning just described, see \cite{KS}.}. Specifically, space-time
independent fields are amenable to a study within this framework. 

The task of constructing an effective 
Lagrangian can be considerably simplified 
by going over to linear combinations of original 
operators\footnote{
In this section we change the normalization of the gluon field 
in comparison to that used in Sect.3 by the rescaling $ A_{\mu} 
\rightarrow (1/g) A_{\mu} $, and use the one-loop 
$ \beta $-function.} 
which enter relations (\ref{3}), (\ref{5}) :
\beq
\label{9}
H  = \frac{b}{64 \pi^2} \left( -
G^2  + i \, \frac{2}{ b \xi} \ 
G \tilde{G} \right) \; , \; \bar{H}   
= \frac{b}{64 \pi^2} \left( - 
G^2  - i \, \frac{2}{b \xi} 
G \tilde{G} \right) \; .
\eeq  
In terms of these combinations, the low energy theorems 
for renormalized zero momentum Green function, Eqs. (\ref{3}) 
and (\ref{5}), take particularly simple forms (for an arbitrary
value of the vacuum angle $ \theta $):
\bea
\label{10}
\lim_{q 
\rightarrow 0 } \, i \int dx 
e^{iqx} \lo T \{ H(x) \; H(0) \} \ro &=& - 4 \la H \ra \; , 
\nonumber \\
\lim_{ q \rightarrow 
0} \,  i \int dx 
e^{iqx} \lo T \{ \bar{H}(x) \; \bar{H}(0) \} \ro &=& - 4 
\la \bar{H} \ra \; , \\ 
\lim_{ q \rightarrow 
0} \,  i \int dx 
e^{iqx} \lo T \{ \bar{H}(x) \; H(0) \} \ro &=&  0 \; . \nonumber 
\eea
It is easy to check that the decoupling of the fields $ H $ 
and $ \bar{H} $ 
holds for arbitrary n-point functions 
of $ H $, $ \bar{H} $. This circumstance makes it particularly 
convenient to work with fields (\ref{9}). 

We now wish to construct an effective low energy Lagrangian 
reproducing at the 
tree level all Ward identities (low energy theorems) 
for the composite fields $ H , \bar{H} $, such as 
Eqs.(\ref{10}) and their n-point generalizations\footnote{
For the case of one real ``dilaton" fields $ \sigma = - b \alpha_s/
(8 \pi) \, G^2 $, a similar problem of constructing an effective 
Lagrangian was solved long ago by Schechter \cite{Sch}, and Migdal
and Shifman \cite{MS} (see also \cite{CS}). Our derivation below
is akin to the one suggested 
by Cornwall and Soni in \cite{CS}.}. To this end, we 
consider the generating functional of connected Green 
functions with the space-time independent sources $
J , \bar{J} $
\bea
\label{10a}
\exp [ i W(J, \bar{J}) ] = \sum_{n} \int DA \exp \left[
- \frac{i}{4 g^2} \int dx \, G^2 + i \frac{ \theta + 2 \pi n}{32 \pi^2}
\int  dx \, G \tilde{G} \right. \nonumber \\ 
\left. + i J \int dx \, H + i \bar{J} \int dx \, \bar{H}
\right] \; .
\eea  
Note the (somewhat
unconventional) summation over all integer numbers $ n $ 
in Eq.(\ref{10a}), which is analogous to the definition (\ref{a3})
for the Schwinger model.
 This prescription automatically ensures the 
$ 2 \pi $ periodicity in $ \theta $ and quantization of the 
topological charge, and is completely equivalent to the way the 
vacuum angle $ \theta $ has initially appeared in YM theory
\cite{Jac}. The above form of introducing the $ \theta $ angle in the 
path integral will help us to understand how the $ \theta $ parameter
should be installed in the effective Lagrangian formalism.

We next define the effective zero momentum fields (here and 
in what follows
$ \int dx = V $ is a total 4-volume)
\beq
\label{10b}
\int dx \, h = \frac{ \partial W}{ \partial J} \; \; , \; \; 
\int dx \, \bar{h} = \frac{ \partial W}{ \partial \bar{J} } \; ,
\eeq
satisfying the equations
\beq
\label{10bb}
\int dx \, h = \la \int dx \, H \ra \; \; , \; \; 
\int dx \, \bar{h} =  \la \int dx \, \bar{H} \ra \; .
\eeq
The effective action $ \Gamma( h ,\bar{h}) $
is now introduced as the Legendre transform of the 
generating functional $ W(J, \bar{J}) $:
\beq
\label{10c}
 \Gamma (h ,\bar{h}) =  - 
W(J, \bar{J}) + \int dx \, J h  + \int dx \, \bar{J} \bar{h} \; 
\eeq
which implies
\beq
\label{10d}
\frac{ \partial \Gamma}{ \partial 
\int dx h} =  J \; \; , \; \; 
\frac{ \partial \Gamma}{ \partial 
\int dx \bar{h}} =  
\bar{J} \; .
\eeq
>From the definition (\ref{10a}) and the low energy theorems
(\ref{10}) (and their extensions for arbitrary n-point functions,
see Eq.(\ref{5a})) we obtain
\bea
\label{10e}
\frac{ \partial^{n+1}}{ \partial J^{n+1}} \; W | _{J = \bar{J} = 0}
&=& i^n \, \int dx \, dx_1 \ldots dx_n \la T \left\{ H(x_1) \ldots
H(x_n) H(0) \right\} \ra = (-4)^n \int dx \la H \ra \; \nonumber \\ 
\frac{ \partial^{n+1}}{ \partial \bar{J}^{n+1}} 
\; W | _{J = \bar{J} = 0}
&=& i^n \, \int dx \, dx_1 \ldots dx_n \la T \left\{ 
\bar{H}(x_1) \ldots
\bar{H}(x_n) \bar{H}(0) \right\} \ra = (-4)^n \int dx \la \bar{H} \ra 
\nonumber \\
\frac{ \partial^{k+l}}{ \partial J^{k} \partial \bar{J}^{l} }
\; W | _{J = \bar{J} = 0}
&=& 0 
\eea
(as before, the connected parts of the Green
functions are implied in 
Eq.(\ref{10e})). These equations are solved by  the function 
\beq
\label{10g}
W(J, \bar{J}) = - \frac{1}{4} \int dx \, \la H \ra e^{ -4 J} - 
\frac{1}{4} \int dx \, \la \bar{H} \ra e^{ -4 \bar{J}} \; .
\eeq
Using Eq.(\ref{10b}), we can express the sources $ J , \bar{J} $
in terms of the fields $ h , \bar{h} $ :
\beq
\label{10k}
J = - \frac{1}{4} 
\log \left( \frac{h}{ \la H \ra } \right) \; \; , \; \; 
\bar{J} = - \frac{1}{4} \log 
\left( \frac{\bar{h}}{ \la \bar{H} \ra } \right) \;.
\eeq
Inserting these expressions back to Eq.(\ref{10g}), we obtain
$ W $ as a function of the fields $ h , \bar{h} $. Now the 
definition (\ref{10c}) turns into the differential equation
for the effective potential $ U( h , \bar{h}) = - (1/V) \Gamma(
h , \bar{h}) $ :
\beq
\label{11}
U - h \, \frac{ \partial U}{ \partial h} - 
\bar{ h} \, \frac{ \partial U}{ \partial \bar{h}} = 
- \frac{1}{4} ( h + \bar{h}) \; .
\eeq
This equation is a complex extension of a real differential 
equation for the ``dilaton" effective potential of 
Refs.\cite{Sch,MS}. Here comes 
the aforementioned ambiguity of the effective potential.
Let us compare Eq.(\ref{11}) with the equation for the real
``dilaton" field of \cite{Sch,MS} 
\beq
\label{12}
U(\sigma) - \sigma \frac{ d U}{ d \sigma} = - \frac{1}{4} 
\sigma \; .
\eeq
Eq.(\ref{12}) has the only solution $ U( \sigma) = (1/4) \sigma 
( \log \sigma + const) $. It is the appearance of the 
multi-branched logarithmic function of a complex argument
in Eq.(\ref{11}) that gives rise to the ambiguity which was 
absent in the real equation (\ref{12}). Let us analyse 
the way it appears when Eq.(\ref{11}) is solved. One obvious 
solution of Eq.(\ref{11}) is 
\beq
\label{13}  
U_1 ( h , \bar{h}) = \frac{1}{4}  h \log \frac{ h}{C}
 + \frac{1}{4} \bar{h} \log \frac{ \bar{h}}{ \bar{C} } 
+ D (h - \bar{h})  \; ,
\eeq
where $ C , \bar{C}, D $ are arbitrary complex constants which 
may depend on $ \la H \ra , \la \bar{H} \ra , \theta $.
However, (\ref{13}) is not a single-valued function, and is 
not bounded 
from below.
After the phase rotation $ h \rightarrow h \exp ( 2 \pi i r ) $
with an arbitrary integer $ r $,
the potential (\ref{13}) transforms as 
\beq
\label{14}
U_1 ( h , \bar{h}) \rightarrow U_1 ( h , \bar{h}) + 
\frac{ i \pi r }{ 2} ( h - \bar{h} ) 
\eeq
which is physically unacceptable. 

A way out in this situation is to sum over all integers $ r $ 
in the partition function, as was suggested by Kovner and 
Shifman \cite{KS} in a similar problem arising with 
Veneziano-Yankielowicz effective Lagrangian \cite{VY} for 
SUSY gluodynamics. Yet, in our case this is not the end of the 
story. Indeed we find that there exists another possible 
solution of Eq.(\ref{11}) :
\beq
\label{15}  
U_2 ( h , \bar{h}) = \frac{1}{4 \alpha}  h \log 
\left( \frac{ h}{C} \right)^{\alpha}
 + \frac{1}{4 \alpha} \bar{h} \log 
\left( \frac{ \bar{h}}{ \bar{C} } \right)^{\alpha} 
+ D (h - \bar{h})  \; ,
\eeq
where $ \alpha $ is an arbitrary real number. From now on we 
concentrate on the case when $ \alpha $ is a positive rational
number, $ \alpha = p /q $, where the integers $ p $ and $ q $ are
relatively prime. Using the formula
\beq
\label{16}
\log \, z^{p/q} = \frac{p}{q} \, Log \, z + 2 \pi i ( n + k 
\frac{p}{q}) \; \; , \; \;  n = 0, \pm 1,\ldots \; \; ; \; 
\; k = 0, 1, \ldots , q-1 \; 
\eeq
( here $ Log $ stands for the principal branch of the logarithm),
we see that the second form (\ref{15}) makes no difference in 
comparison with (\ref{13}) when only the principal value of 
the logarithm is considered. However, the theories, described by the
effective potentials (\ref{13}) and (\ref{15}), are different quantum
mechanically as they imply different rules of a global 
quantization for the 
fields $ h , \bar{h} $. This quantization
arises when the single-valuedness of the partition function is 
ensured by a summations over the integers. As will be shown in the 
next section, it is the second choice (\ref{15}) for the 
effective potential that can be made consistent with both the 
$ \theta/ N $ dependence and $ 2 \pi $ periodicity in $ \theta $
when a proper treatment to global quantization constraints and 
the thermodynamic limit is given.

We therefore consider the function 
\beq
\label{17}
U_3 (h ,\bar{h}) = \frac{1}{4}  h \, Log  \, \frac{ h}{C}
 + \frac{1}{4} \bar{h} \, Log \, 
 \frac{ \bar{h}}{ \bar{C} } + \frac{i \pi}{2} \left( k 
+ n \, \frac{q}{p}
\right) \left( h - \bar{h} \right) \; , 
\eeq
which satisfies Eq.(\ref{11}) (i.e.
the Ward identities (\ref{10}) and 
their n-point generalizations) for any values of the 
integers $ n , k $
from the range $ n = 0, \pm 1 \, \ldots ; k = 0,1,\ldots ,
q-1 $. The 
last term in (\ref{17}) is a particular form of the last 
term $ \sim D
$ in Eq.(\ref{15}). It can be seen that arbitrary values of the 
coefficient $ D $ would be uncompatible with the quantization 
rules 
imposed by the summation over the integers $ n , k $ in the 
partition function.

Finally, we have to figure out how the $ \theta $ angle 
should be installed in the effective potential (\ref{17}). An answer 
to this question can be deduced by comparing with Eq.(\ref{10a}).
In the YM partition function, the $ \theta $ angle enters in the 
combination $ \theta + 2 \pi n $, while the summation over the 
integers $ n $ is necessary because of a multi-valuedness of the 
YM action in respect to large gauge transformations. Analogously, 
the last term in Eq.(\ref{17}) is the only one that can accommodate
the $ \theta $ parameter in the same combination $ \theta + 2 \pi n $.
The summation over the integers $ n $ is enforced this time by the 
multi-valuedness of the logarithm in Eq.(\ref{17}). As the presence 
of the $ \theta $ angle is implicit in the constants $ C , \bar{C} $,
we now make it explicit in the above way and finally obtain the
(Minkowsky space) 
improved effective potential $ F (h , \bar{h}) $ by the summation over 
$ n , k $ in the partition function :  
\bea
\label{pot}
e^{- i V F(h,\bar{h}) } &=& \sum_{n = - \infty}^{
 + \infty} \sum_{k=0}^{q-1} \exp \left\{ - \frac{i V}{4}
\left( h \, Log \, \frac{h}{C'} + 
\bar{h} \, Log \, \frac{ \bar{h}}{
\bar{C'}} \right) \right. \nonumber \\ 
&+& \left. i \pi V \left( k + \frac{q}{p} \,  
\frac{ \theta + 2 \pi n}{ 2 \pi} \right) \frac{h - \bar{h}}{
2 i} \right\} \;  ,
\eea
where    
the constants $ C', \bar{C'} $ are independent of $ \theta $ and
can be taken real, $  C' =  \bar{C'} \equiv 2 e E $, 
where $ E $ is some positive constant. The improved 
effective potential $ F(h, \bar{h}) $ is consistent with all 
constraints
imposed by the low energy theorems and, by construction, is a single
valued function possessing the $ 2 \pi $ periodicity in $ \theta $, 
which was present in the initial YM partition function. 
As is seen from Eq.(\ref{pot}), the structure of 
the effective potential $ F $ is such that it contains
both the ``dynamical" and ``topological" parts (the 
first and the second terms in the exponent, respectively). We would 
like to note that Eq.(\ref{pot}) is a direct analog of a similar 
construction for SUSY models \cite{VY,KS}. Namely, 
the ``dynamical" part of the effective potential
is rather similar to Veneziano-Yankielowicz (VY) \cite{VY}
potential $
\sim u^{2/3} \log u $, while the ``topological" part is 
analogous to an improvement of the VY effective potential, 
suggested by Kovner and Shifman \cite{KS}. 
We stress that the improved effective potential
(\ref{pot}) contains more information in comparison to that 
present in the Ward identities (\ref{10}) just due to the 
appearance of this ``topological" part in Eq.(\ref{pot}).
Without this term Eq.(\ref{pot}) would merely be a kinematical
reformulation of the content of the Ward identities (\ref{10}).
As will be shown in the next section, this improvement 
of the effective potential turns out crucial for unravelling 
the correct periodicity in $ \theta $ in YM theory. 
 
\section{Minimization of effective potential}

In this section a ground state of the dual low energy theory
will be determined by a minimization of the improved effective
potential (IEP) $ F(h,\bar{h}) $ given by Eq.(\ref{pot}).
Our purpose is to find the $ \theta $ dependence of the 
vacuum energy which is  defined as a minimum of IEP $ F(h,\bar{h}) $.
In this calculation the total space-time 4-volume will be kept 
finite, while a transition to the thermodynamic limit $ V \rightarrow
\infty $ will be performed at the very end. 

We start with introducing the ``physical" real fields $ \rho ,\eta $
defined by the relations
\beq
\label{29}
h = 2 E \, e^{ \rho + i \eta} \; \; , \; \; 
\bar{h} = 2 E \, e^{ \rho - i \eta } \; .
\eeq
(This definition implies $ F(\eta + 2 \pi ) = F(\eta) $. As 
will be seen, this condition of single-valuedness of the $ \eta $
field is satisfied with the substitution (\ref{29}).) In these 
variables, the ``dynamical" part of Eq.(\ref{pot}) 
can be written as follows:
\beq
\label{30}
- \frac{iV}{4} \left( h \, Log \frac{h}{2 e E} + 
 \bar{h} \, Log \frac{ \bar{h}}{ 2 e E} \right) =
- i V E \, e^{ \rho} \left[ ( \rho -1) \cos \eta - \eta \sin \eta 
\right] \; .
\eeq
The summation over the integers $ n $ in Eq.(\ref{pot}) enforces 
the quantization rule due to the Poisson formula 
\beq
\label{31} 
\sum_{n} \exp \left(
2 \pi i n  \, \frac{q}{p} \, V \, \frac{h - \bar{h}}{ 4 i} 
\right) = \sum_{m} \delta \left( \frac{q}{p}
\, V E e^{ \rho} \, \sin \eta - m \right) \; ,
\eeq 
which reflects the quantization of the topological charge
in the original theory.
Therefore, when the constraint (\ref{31}) is imposed,
Eq.(\ref{30}) can be written as 
\beq
\label{32}
- \frac{iV}{4} \left( h \, Log \frac{h}{2 e E} + 
 \bar{h} \, Log \frac{ \bar{h}}{ 2 e E} \right) =
- i V E \, e^{ \rho} \, ( \rho -1) \cos \eta 
+ i m \, \frac{p}{q} \, \eta \; .
\eeq
Using (\ref{31}),(\ref{32}), we put Eq.(\ref{pot}) in the form
\beq
\label{33}
e^{ - i V F} = \sum_{m = - \infty
}^{ + \infty} \sum_{k=0}^{q-1}
 \delta( V E \, \frac{q}{p} \, e^{ \rho} 
\sin \eta - m) \, \exp \left[ - i V E \, e^{ \rho} ( \rho -1)
\cos \eta 
+ i m \left( \theta_k +  \frac{p}{q} \, \eta 
\right) \right] 
\eeq
where we denoted
\beq
\label{34}
\theta_k \equiv \theta + 2 \pi \, \frac{p}{q} \, k \; .
\eeq
To resolve the constraint imposed by the presence 
of $ \delta $-function in Eqs.(\ref{31}),(\ref{33}),   
we introduce the new field $ M $ by the formula
\beq
\label{35}
 \delta( V E \, \frac{q}{p} \, e^{ \rho} 
\sin \eta - m) \propto \int D \, M \; \exp \left( i M 
V E \, e^{ \rho } \sin \eta - i M \, \frac{p}{q} \, m \right) \; 
\eeq
Going over to Euclidean space\footnote{This is not really necessary.
All formulas below can be worked out in Minkowsky space as well.}
 by the substitution 
$ i V \rightarrow V $,
we obtain from Eqs.(\ref{33}),(\ref{35}) 
\bea
\label{36}
F( \rho, \eta, M) = - \frac{1}{V} \log \left\{ \sum_{m =
- \infty}^{+ \infty} \sum_{k=0}^{ q-1} \exp 
\left[
- V E e^{ \rho} \left\{ ( \rho - 1) \cos \eta - M \sin \eta \right\} 
  \right. \right.\nonumber \\ 
+  \left. \left.
i m  \left( \theta + 2 \pi k \, \frac{p}{q} + \frac{p}{q} \, \eta - 
\frac{p}{q} \, M \right) -  \varepsilon \, 
\frac{m^2}{ VE} \right] 
\right\} \; .
\eea
Here we introduced the last term to regularize the infinite sum over  
the integers $ m $. The limit $ \varepsilon \rightarrow 0 $ will be 
carried out at the end, but before taking the thermodynamic limit 
$ V \rightarrow \infty $. Note that Eq.(\ref{36}) satisfies the 
condition $ F (\eta + 2 \pi) = F (\eta) $ which should hold as 
long as $ \eta $ is an angle variable. We also note that the 
periodicity in $ \theta $ with period $ 2 \pi $ is explicit in
Eq.(\ref{36}).

Proceeding as was done for the Schwinger model, 
to discuss the thermodynamic limit $ V \rightarrow \infty $ we 
use the identity (\ref{a7}) and transform Eq.(\ref{36})
into its dual form
\bea
\label{38}
F( \rho, \eta, M) = - \frac{1}{V} \log \left\{ \sum_{n =
- \infty}^{ + \infty} \sum_{k=0}^{ q-1} \exp 
\left[
- V E e^{ \rho} \left\{ ( \rho - 1) \cos \eta -
 M \sin \eta \right\} 
  \right. \right. \nonumber \\ 
-  \left. \left.
\frac{VE}{ 4 \varepsilon} 
\left( \theta  +  2 \pi k \, \frac{p}{q} + \frac{p}{q} \, 
\eta - 
\frac{p}{q} \, M - 2 \pi n \right)^2  \; \right] \right\} \; ,
\eea
where we have omitted an irrelevant infinite factor  
$ \sim \varepsilon^{-1/2} $
in front
of the sum. 
Eq.(\ref{38}) is the final form of the improved effective potential
$ F $, which represents the YM analog of 
Eq.(\ref{a8}) for the Schwinger 
model. To discuss the vacuum properties, the function $ F $ 
should be minimized in respect to the three variables 
$ \rho , \eta $ and $ M $. In spite of the frightening form
of this function, its extrema can be readily found using the following 
simple trick. As at the extremum points all partial derivatives of the 
function $ F $ vanish, we first consider their linear combination
in which the sum over $ n , k $ cancels out. We thus arrive at the 
equations
\bea
\label{39}
\frac{ \partial F}{ \partial \rho} =  E e^{ \rho}
( \rho \cos \eta - M \sin \eta ) &=& 0 \nonumber \\
\frac{ \partial F}{ \partial \eta} + 
\frac{ \partial F}{ \partial M} = - E e^{ \rho} (
\rho \sin \eta + M \cos \eta ) &=& 0 \; ,
\eea
which is equivalent to $ \rho^2 + M^2 = 0 $. Therefore,
these equations have the only solution 
\beq
\label{40}
\la \rho \ra = 0 \; \; , \; \; \la M \ra = 0 \; , 
\eeq
while the minimum value of the 
angular field $ \eta $ is left arbitrary by them.
The latter can now be found from either of the 
constraints $ 
\partial F / \partial \eta = 0 $ or 
$ \partial F / \partial M = 0 $, which become identical for $ 
\la \rho \ra  = \la M \ra = 0 $. The resulting equation 
reads\footnote{Note a remarkable similarity between Eq.(\ref{41}) 
and
the equation $ \mu_{i}^2 \sin \phi_i = (a/N) ( \theta - \sum
\phi_{j} ) $ (where $ \mu_{i} $ and $ \phi_{i} $ are the masses 
and phases of goldstone fields, respectively), obtained by Witten
\cite{Wit2} as a minimization condition for the effective chiral
Lagrangian for QCD. The limit $ \varepsilon \rightarrow 0 $ 
in Eq.(\ref{41}) is analogous to the chiral limit $ \mu_{i}^2 
\rightarrow 0 $ in this equation.}
\bea
\label{41} 
\sum_{n =
- \infty}^{+ \infty} \sum_{k=0}^{q-1} 
\left( \theta + 2 \pi k \, \frac{p}{q} - 
2 \pi n + \frac{p}{q} \, 
\eta +  2 \varepsilon \, \frac{q}{p} \sin \eta \right) \nonumber \\
\times \exp \left\{ V E \cos \eta  
 - \frac{VE }{ 4 \varepsilon}
\left( \theta + 2 \pi k \, \frac{p}{q}  + \frac{p}{q} \, 
\eta  - 2 \pi n \right)^2 \right\} = 0 \; ,
\eea
in which we have to take the limit $ \varepsilon \rightarrow 0 $ 
at a fixed 4-volume $ V $. 

One can see that non-trivial solutions of Eq.(\ref{41}) at 
$ \varepsilon \rightarrow 0 $ are given by 
\beq
\label{42}
\la \eta \ra_l = - \frac{q}{p} \, \theta + \frac{2 \pi}{ p}
\, l + 2 \pi r \; \; , \; \; l = 0,1, \ldots , p-1 \; \; ; 
\; \; r = 0, \pm 1, \ldots
\eeq
Eq.(\ref{42}) shows that there are $ p $ 
physically distinct solutions, while 
the series over the integers $ r $ in Eq.(\ref{42}) simply reflects the
angular character of the $ \eta $ variable, 
and is thus irrelevant. By the substitution
of Eq.(\ref{42}) back to Eq.(\ref{38}) we obtain the energy spectrum
for the finite volume theory:
\beq
\label{43}
E_l \equiv F \left( \rho = M = 0 , \eta = \la \eta \ra_l \right) = 
- E \cos \la \eta \ra_l = 
- E \cos \left( - \frac{q}{p} \theta + \frac{2 \pi}{ p} \,  
l \right)
\;  .
\eeq
Thus, we have found that the improved effective potential (\ref{38})
has not one, but rather $ p $ physically different local extrema, 
when we look at the theory in the finite volume. The fact that the 
states (\ref{42}) have different energies for generic values of 
$ \theta $
is very important. This is where we find an essential difference of 
non-supersymmetric gluodynamics from its supersymmetric extension.
In the latter case, there are $ N_c $ degenerate states which all 
survive the infinite volume limit, and correspond to the physical
$ Z_{2N_c} $ symmetry of SUSY YM theory \cite{SV,KS,SUSY}. The 
absence of 
degeneracy between the states (\ref{42}) is therefore very natural,
as there are no discrete symmetries for 
non-supersymmetric gluodynamics in the thermodynamic limit
$ V \rightarrow \infty $ where 
we should stay along with just one true
vacuum. Yet, as we will see in a moment, retaining
the whole set of local extrema (\ref{42}) is important
to recover the correct periodicity
in $ \theta $ in the limit $ V \rightarrow \infty $.

The remarkable fact about the extrema (\ref{42}) is that they are
related to each other by a cyclic permutation under the shift 
$ \theta \rightarrow \theta + 2 \pi $. The physics is perfectly 
periodic in $ \theta $ with period
$ 2 \pi $, as the minima $ \la \eta \ra_l $, 
interchanging under the shift 
$ \theta \rightarrow \theta + 2 \pi $,
 can be just re-labeled without altering 
anything.
One of the minima 
always has a lowest energy. For example, if  
$ 0 \leq \theta < \pi/q $, it is the $ l = 0 $ solution 
in Eq.(\ref{42}). 
At the same time, we observe  
level crossing with a two-fold degeneracy at 
certain values of $ \theta $. One series of the 
level crossing points is given by $ \theta = 
\pi \; (mod \, 
2 \pi ) $, irrespective of the values of the integers $ p ,q $. For
example,
at the first point $ \theta =  \pi $ in this series, the $ l = 0 $
and $ l = q $ solutions have the same energy $ - E \cos( \pi q /p) $.
This is the same series of level crossing points as was found for 
the Schwinger model. The difference from the Schwinger model is that
now these values of $ \theta $ do not correspond to level crossing
of {\it lowest energy} states among the set (\ref{42}). Instead,
this happens for another 
series 
of level crossing points in $ \theta $, which is different from 
the previous one as long as $ q \neq 1 $. As can be seen from 
Eq.(\ref{43}), it is the points $ \theta_{k} = (2 k + 1) \pi/q $ , 
$ k = 0, 1, \ldots , p-1 $ where the lowest energy state is 
changing from the $ k $th to the $(k+1) $th branch in 
the set (\ref{42}).    

Let us now see what happens when the thermodynamic 
limit $ V \rightarrow 
\infty $ is taken. The key observation is that a lowest energy 
state, which is the only one that should
be retained in the limit $  V \rightarrow 
\infty $ according
to our convention
(\ref{e}), corresponds to different minima from the set (\ref{42}),
depending on an interval of variation of the vacuum  $ \theta $
angle. 
Thus, to perform the thermodynamic limit, we should 
first fix an interval of $ \theta $ (say, $ 0 \leq \theta < \pi/q $),
and only then select the state of lowest energy among 
the set (\ref{42}). This solution will be the one corresponding 
to the single vacuum state in the limit $ V \rightarrow \infty $, for
all values of $ \theta $ from this interval. 
 This procedure can be described  
by the formula 
\beq
\label{44}
F_{min} =  - \lim_{V \rightarrow 
\infty} \; \frac{1}{V} \log \left\{ 
\frac{ \sum_{l} \exp \left[ 
V E \cos \left( - \frac{q}{p} \, \theta + \frac{2 \pi}{p}
\, l \right) \right] }{ 1 + \delta_{ \theta 
- \pi/q, 0}} \right\}  \; \; , \; \; l= 0,1, \ldots , p-1 
\eeq
(here $ \delta_{ \theta - \pi/q, 0} $ is the Kronecker symbol, 
equal to 1 if  $ \theta =  (2 k + 1) \pi/q  $ or 0 otherwise). 
The multiplier $ 1/( 1 +  
\delta_{ \theta - \pi/q, 0}) $ accounts for the two-fold 
vacuum degeneracy at the points $ \theta =  (2 k + 1) \pi/q \; , 
\; k = 0,1, \ldots , p-1  
 $. We note that Eq.(\ref{44})
is perfectly periodic in $ \theta $ with period $ 2 \pi $.

Eq.(\ref{44}) shows that in the limit $ V \rightarrow \infty $
cusp singularities occur at the values $  \theta = (2 k + 1) \pi/q  
$, where the lowest energy vacuum state switches from one
analytic branch to another one, much as it occurs 
in the Schwinger model. 
One should note that there is no
physical jump at $ \theta = \pi/q $.
It is rather re-labelling of a lowest 
energy state. The first derivative of the vacuum 
energy, which is proportional to the topological
density condensate, is two-valued at these points.
This means that whenever $ \theta =  (2 k + 1) \pi/q $, we stay  
with two degenerate vacua in the thermodynamic
limit (Dashen phenomenon \cite{Dash}, see below). 
This picture of the singularity structure in $ \theta $ resembles the
one found for the lattice $Z_p $ model in 4D \cite{Car}.

If, on the other hand, the thermodynamic
limit is performed for a fixed value of $ \theta $,
any
information on other states is completely lost in Eq.(\ref{44}).
Correspondingly, the $ 2 \pi $ periodicity in $ \theta $ is 
also lost in infinite volume formulae. 
We have 
no chance to know about additional states when we work in the 
infinite volume limit from the very beginning. As a result,
usual $ V = \infty $ formulae become blind to the very existence 
of a whole set of different vacua,
which is just responsible
for restoration of the $ 2 \pi $ periodicity in $ \theta $. 
Instead, formulae
corresponding to the formal limit $ V = \infty $ look as suggesting
a ``wrong" (different from $ 2 \pi $) periodicity in $ \theta $ ,
see e.g. Eqs.(\ref{7},\ref{8}). Now we know that this procedure 
of the shift $ \theta \rightarrow 
\theta + 2 \pi $ in the $ V = \infty $ formulae is simply misleading
as it is equivalent to going along a
single analytic solution of the 
minimization equation (\ref{41}), which does not corresponds to 
a lowest energy state for a shifted value of $ \theta $. 
Comparing 
Eqs.(\ref{7}) and (\ref{43}),
we see that the former may well describe the $ \theta $ dependence 
in the physical limit $ V \rightarrow \infty $ for small values
$ \theta < \pi/q $. To this end, we should set the ratio $ q/p $, which 
so far was arbitrary, to the value $ q/p = 2 \xi $
(and take $ E = - E_v $). At the same time,
analyticity in $ \theta $ of each separate branch (\ref{43}) shows 
that the procedure of a formal re-summation of the infinite Taylor 
series for small $ \theta < \pi/q $, which has led to Eq.(\ref{7}) 
\cite{6}, is legitimate. We therefore conclude that 
Eq.(\ref{7}), which should be understood as standing for 
$ \theta < \pi/q $, is not in conflict with general principles
of periodicity and analyticity in $ \theta $ for any rational 
value of the parameter $ \xi $, including both particular
choices $ \xi = 2/b $ \cite{NSVZ} or  
 $ \xi = 4/(3 b) $ \cite{6}. 

Although the problem of fixing the correct value of the parameter
$ \xi $ is beyond the scope of this paper, we can not refrain
from pausing for a few comments on these matters\footnote{
The arguments discussed below are due to 
A. Vainshtein \cite{V}, to whom we are indebted 
for sharing with us his insight.}. The parameter $ \xi $ is 
related to a number of different sectors of the theory, which 
are disconnected due to the superselection
rule. One could think that this number 
of sectors is proportional
to $ b $, as the formulae $ \xi = 4/(3b) $ \cite{6} or 
$ \xi = 2/b $ \cite{NSVZ} suggest. However, the analysis of SUSY
theories shows that it might not literally be the case. If 
in SQCD we 
change a number of flavors $ N_f $ keeping a number of colors
$ N_c $ unchanged, the number of sector remains the same 
and strictly equals $ N_c $
(and not just proportional to $ N_c $), 
though the $ \beta $-function $ b \sim 3 N_c 
- N_f $ changes.  As a result, the angle $ \theta $ enters physics 
in the combination $ \theta/ N_c $ for arbitrary $ N_f $.
Another argument comes from the 
analysis of softly broken SUSY gluodynamics
where the gluino is given a small mass $ m $. As discussed by Shifman
\cite{SUSY}, the situation is under control as long as $ m $ is small,
and the number of different sectors remains the same $ N_c $ (though
the degeneracy between them is lifted). 
On the other hand, the case 
of pure YM theory corresponds to the limit $ m \rightarrow \infty $
(which means physically $ m \gg \Lambda $). 
If the number of sectors $ N_c $ for small 
$ m $ does not discontinuously changes when $ m $ 
becomes large, $ m \simeq 
\Lambda $, we end up with $ N_c $ 
different
sectors for usual YM theory.
   
To summarize, different lines of reasoning lead to different values 
of $ \xi $ (though they all imply $ \xi \sim N_{c}^{-1} $),
 and correspondingly to different values for 
 a number of sectors in pure 
YM theory. However, irrespective of this particular 
number (we only assume it be a  
rational), we know that there is only one true vacuum in the 
thermodynamic limit 
$ V \rightarrow \infty $ and that the period of the  $ \theta $ 
dependence is always $ 2 \pi $. 

 After this digression we wish to discuss relations for 
the topological density condensate. 
Namely, we would like to see whether Eq.(\ref{8}), obtained 
by a direct evaluation of correlation functions in YM theory,
is consistent with results of this section. 
It is easy to see that it is indeed the case.
Differentiating Eq.(\ref{43}) in respect to $ \theta $, we obtain
\beq
\label{45}
- \frac{ \partial E_l}{ \partial \theta} = 
\frac{q}{p} \, E \sin \left( - \frac{q}{p} \, \theta + \frac{2
\pi}{ p} \, l \right) \; .
\eeq
As only the $ l = 0 $ term survives the $ V \rightarrow \infty $ 
limit for $ 0 \leq \theta < \pi/q $, we obtain, in agreement with
Eq.(\ref{8})
\beq
\label{46}
\frac{1}{32 \pi^2} \la G \tilde{G} \ra = - 
\frac{q}{p} \, E \sin \left( \frac{q}{p} \, \theta
 \right) \; .
\eeq
Similarly to what occurs 
in the Schwinger model, for the special case $ \theta = \pi/q $
we stay in the limit $ V \rightarrow \infty $ with two 
degenerate vacua which are distinguished by the sign of the 
topological density condensate:
\bea
\label{47}
| 1 \ra & \equiv &  | \eta_{l=0} = -  \frac{\pi}{p} \ra \; \; , \; \; 
\frac{1}{32 \pi^2} \la G \tilde{G} \ra_1 = -   
\frac{q}{p} \, E \sin \left( \frac{\pi}{p} 
 \right)  \nonumber \\
| 2 \ra & \equiv & | \eta_{l=1} =  \frac{\pi}{p} \ra \; \; , \; \; 
\frac{1}{32 \pi^2} \la G \tilde{G} \ra_2 = +    
\frac{q}{p} \, E \sin \left( \frac{\pi}{p}
 \right) 
\eea
As a CP transformation reverses the sign of $ \theta $, it 
exchanges the vacua $ | 1 \ra $ and $ | 2 \ra $:
$ CP |1 \ra = | 2 \ra $ , $  CP |2 \ra = | 1 \ra $ . Therefore, 
the CP symmetry is broken at $ \theta = \pi/q $. 
A similar phenomenon of vacuum doubling occurs for any point of the 
form $ \theta_k = k \pi/q $, $ k = 1,2,3, \ldots $. For example,
the $ l = 0 $ and $ l = q $ states are analogously related by a 
CP transformation at $ \theta = \pi $. The reason we 
concentrate on the level crossing point $ \theta = \pi/q $ is that 
at this value of $ \theta $ the true vacuum (lowest energy state) 
switches from the $ l = 0 $ to the $ l = 1 $ branch, while at 
$ \theta = \pi $ some excited states cross in energy. Therefore,
in the sense of the lowest energy state among the set (\ref{42}), 
the value $ \theta = \pi $ corresponds to a regular, CP conserving
point. 

Finally, we would like to briefly describe the case of YM 
theory with an arbitrary 
(orthogonal, exceptional etc.) gauge group $ G $ instead of 
the unitary SU(N) that was discussed so far. For such a gauge group
the second Casimir constant $ C_2 (G) $ and 
the $ \beta $-function would be 
different. Therefore, 
 the only difference from
the previous analysis in this case would be different values of 
the integers $ p $ and $ q $, while 
the pattern of $ \theta $ 
dependence would remain the same. Thus, the mechanism suggested
in this paper seems to be valid for any gauge group.

\section{Conclusions}

The most important results of the present analysis are
the following:\\ 
1. We have demonstrated that physics is periodic  
in $\theta$ with period 
  $2\pi$ for an arbitrary gauge group. This behavior
follows from our definition of the partition function
for both the original and effective Lagrangian formulations
of the theory, 
where the summation over all branches
 of a  multivalued action is imposed. In the effective Lagrangian 
framework, this prescription is necessary for a single-valuedness
and boundness from below of an effective potential.

2.The periodicity
in $ \theta $ with period $ 2 \pi $  is perfectly 
compatible with the $\theta/N$ dependence
found in a number of models at small $\theta$. 
The correct periodicity in $ \theta $ is recovered when
a whole set of different branches is taken into account.
The standard definition of the thermodynamic
limit selects only a lowest energy state among this set.
As a result, the thermodynamic 
limit and
the shift $\theta\rightarrow\theta+2\pi$ do not commute (in the sense
explained in the Introduction).
  
3. For generic values of $\theta$, there is one and only 
one vacuum state  in the 
thermodynamic  limit. For $\theta= \pi/q $ there are  exactly
two degenerate states, which are distinguished by the sign
of the topological density condensate.
 
We would like to end up
with some speculations.
We emphasize again that 
    $\theta= \pi/q $ is a very special point because of the  
  vacuum degeneracy which does not follow from  
any obvious symmetry of the original Lagrangian.
This degeneracy may imply the existence of domain walls 
in the theory at $\theta= \pi/q $, 
which are static field configurations
depending only on one spatial coordinate.
An effective potential describing domain walls could be 
obtained from Eq.(\ref{38}) by freezing the $ \rho $ and 
$ M $ fields. Such a potential is a complicated function
of the $ \eta $ field which, however, reduces to the 
standard Sine-Gordon form near the points (\ref{42}), see
Eq.(\ref{43}). Provided a kinetic term (we recall that 
kinetic terms are not fixed by the Ward identities) is added,
the theory could sustain domain wall configurations. 
The existence of these solutions in gauge theories
could have interesting 
consequences for cosmology and particle physics.

We would also like to speculate that the above $ ``p"$ 
non-equivalent states, 
could be really observed in some 
nonequilibrium high energy processes with
a finite geometry, where the superselection
rule can not be applied (a similar comment was made 
by Shifman \cite{SUSY}).
Such a situation could be realized e.g.
in nuclear-nuclear collisions, where the appearance of droplets
of a ``false vacuum" would be similar to the production of 
droplets 
of disoriented chiral condensate, see e.g. \cite{Bj} for a 
review. 
Instead of an arbitrary direction of the chiral condensate
for the latter, in the former case we would deal with $ ``p" $
different values of the topological density condensate 
$ \la G \tilde{G} \ra $.
One expects that this phenomenon, if exists, 
should be related to the physics of 
the $ \eta' $ meson and CP violation.
Yet, it is not known at the moment how to formulate this 
problem in an appropriate way.

The inclusion of the light quarks into the effective Lagrangian
framework and the resulting picture of the $ \theta $ dependence 
in QCD will be discussed in \cite{QCD}.  

\section*{Acknowledgements}

We are indebted to Arkady Vainshtein and Michael Shifman 
for their criticism of a first version of this work,
where the integer-valued Lagrange multiplier was not introduced,
which resulted in the wrong conclusion on the disappearance
of physical $\theta$ 
dependence in gluodynamics.
We are thankful to them for 
comments which motivated this study.
We would like to thank David Gross for his interest  and 
discussions.

\clearpage
\appendix
\def\theequation{\thesection.\arabic{equation}}

\section*{Appendix}

\def\thesection{A}
\setcounter{equation}{0}
 
The purpose of this appendix is to discuss in somewhat 
more detail the derivation of  
the low energy theorems (\ref{3}) and (\ref{5}) and, in particular,
a procedure of  ultra-violet regularization which was implied 
in Eqs.(\ref{3}) and (\ref{5}).
We start with the NSVZ low energy theorem \cite{NSVZ}, Eq.(\ref{3}), 
which is here repeated for convenience:

\beq
\label{b3}
\lim_{q \rightarrow 0 } \; i 
\int dx \, e^{iqx} \lo T 
\{ \frac{\beta(\alpha_s)}{4 \alpha_s} G^2 (x)  \,  
 \frac{\beta(\alpha_s)}{4 \alpha_s} G^2 (0) \} \ro  = 
-4  \la \frac{
\beta(\alpha_s)}{ 4 \alpha_s} G^2 \ra \; ,
\eeq
where $ \beta(\alpha_s) = 
- b \alpha_{s}^2 /(2 \pi) + O(\alpha_{s}^3 ) $ is the
Gell-Mann - Low $ \beta $-function for YM theory with
 $ b = (11/3) N_c $, and $ N_c $ is the number of colors.
 
The low energy theorem was obtained 
in \cite{NSVZ} using the one-loop $ \beta $-function with
a particular attention to a regularization of ultra-violet
(UV) divergences, arising in the two-point 
function (\ref{b3}), within Pauli-Villars 
procedure. In this derivation Dyson type of the T-product symbol
was implied in Eq.(\ref{b3}). 
It was shown that quadratically divergent UV 
contributions cancel out identically in both sides of Eq.(\ref{b3}).
This implies that perturbative contributions should always
be subtracted in vacuum condensates such as 
$ \la \beta(\alpha_s)/
(4 \alpha_s) G^2 \ra $ in Eq.(\ref{b3}). Once this rule 
is accepted,
the dependence
of any (nonperturbative) 
condensate $ \la O \ra_{NP} $ of dimension $ d $  
on the bare coupling constant $ g_0 $ (normalized at the cut-off
scale $ M_R $) is fixed by the dimensional transmutation 
formula:
\beq
\label{b4}
\la O \ra_{NP} = const \; \left[ M_R \exp \left(- 
\frac{8 \pi^2}{b
g_0^2} \right) \right]^d  \; ,
\eeq
and the derivation of the NSVZ theorem proceeds as described in Sect.3,
where the path integral definition of correlation functions is used.
The latter implies Wick type of the T-product symbol. This definition 
of zero momentum correlation functions (\ref{5a}) automatically ensures
the same type of renormalization for all such functions, 
which is fixed by a rule of subtracting perturbative UV divergent 
contributions to the conformal anomaly (with e.g. Pauli-Villars
regularization). The latter procedure thus defines a nonperturbative 
gluon condensate in the $ \theta $-vacuum $ \la g^2 G^2 \ra_{\theta}
$, i.e. a nonperturbative 
part of the conformal anomaly
calculated for given $ \theta $. 
Its dependence on $ 1/g_{0}^2 $ is given by Eq.(\ref{b4}). Zero 
momentum correlation functions of the 
operator $ g^2 G^2 $ are obtained   
by the differentiation of the nonperturbative part
of the partition function $ \log (Z/Z_{PT}) $ (
$ Z_{PT} $ stands for a perturbatively defined partition 
function which does not depend on $ \theta $), where 
\beq
\label{z}
Z (\theta ) =  Z_{PT} \exp \left\{ - i V E_{v}(\theta) \right\}
 = Z_{PT} \, \exp \left\{ - i V  \lo 
- \frac{ b \alpha_s}{ 32 \pi}
G^2 \ro_{\theta} \right\} \; ,
\eeq 
in respect to $ 1/g_{0}^{2} $. Note that the factor $ \partial \, 
\log Z_{PT} / \partial \, (g_{0}^{-2} ) $ corresponds to the 
correlation function (\ref{b3}) in perturbation theory.

It is a subtraction of perturbative contributions in vacuum
condensates (\ref{b4}) that we here would like to comment upon.
Technically, 
this prescription can be thought of as 
the requirement of absence of regular powers of the coupling 
constant 
$ \alpha_s $ in vacuum condensates $ \la O \ra_{NP}
 $ to any finite order in $ \alpha_s $.
For two-dimensional models, it has been shown \cite{sigma} that 
the definition of vacuum condensates via the path integral 
automatically 
nullifies perturbative contributions to the condensates. Moreover, this 
procedure gives results identical to a point-splitting regularization
with Dyson type of the T-product. One can notice that    
in four dimensions the separation of 
genuinely perturbative and nonperturbative contributions to physical
quantities is ambiguous as it depends on a definition of the sum of
perturbative series. Still, there is nothing wrong with
the requirement that the ``nonperturbative" condensates contain
the coupling constant $ g_{0}^2 $ as in Eq.(\ref{b4}), while 
the regular powers of $ g_{0}^2 $ are absent. The difference 
between different regularization schemes is reduced in this case to 
possible finite renormalizations (different numerical values)
of the vacuum condensates. Such a choice of the nonperturbative
gluon condensate in Eq.(\ref{z}) is the only ambiguity for all
multi-point correlation functions (\ref{5a}).
 
 Next we would like to discuss zero momentum 
correlation functions of the topological density operator.
With Wick definition of the T-product we obtain from Eq.(\ref{z})
\beq
\label{b7}
\lim_{ q \rightarrow 0} \; 
i \int dx \, e^{iqx} \lo T \left\{ \frac{\alpha_s}{8 \pi} 
G \tilde{G} (x)  \, 
\frac{\alpha_s}{8 \pi} G \tilde{G} (0) \right\}  \ro = 
- \frac{ \partial^2 E_{v}(\theta)}{\partial \theta^2} = 
- \frac{1}{4} \frac{ \partial^2}{ \partial \theta^2} \, \la 
\sigma \ra_{\theta} \; .   
\eeq
(Eq.(\ref{b7}) is written for Minkowsky space, and $ \sigma $
stands for the trace of the energy-momentum tensor.) 
A few comments on Eq.(\ref{b7}) are in order. 
We note that the two definition (through 
Dyson or Wick T-products) are equivalent for the 
correlation function (\ref{b3}). For correlation 
functions of the  topological density $ Q $, this is no longer 
the case. Witten-Veneziano construction \cite{Wit}
specifically implies \cite{DE}
Wick T-product in the two-point function (\ref{b7}). This can be 
seen both from the definition of zero momentum correlation functions
of the operator $ Q $ used in \cite{Wit}, and from the fact 
the a gauge non-invariant axial ghost (Veneziano ghost pole)
can not appear in Dyson T-product which is related to contributions
of intermediate gauge-invariant states only. This is why Wick type 
of the T-product was used in Eq.(\ref{b7}).
With this definition the two-point function (\ref{b7}) 
does not contain UV divergences which are present in $ Z_{PT} $
and drop out after the differentiating of $ \log Z $ in respect to 
$ \theta $.
We note that an attempt to calculate the correlation function
(\ref{b7}) using Dyson T-product (and adding corresponding
contact terms) 
would face a problem due to the 
fact that, 
in contrast to the 
trace of the energy-momentum tensor, the topological density
operator in pure YM theory is not seen \cite{VZ}
to be related to any quantity conserved at the 
classical level. Therefore, the canonical methods with 
Pauli-Villars regularization procedure  
used in \cite{NSVZ} would  
apparently be not 
applicable in this case.

We further require that the $ \theta $ dependence in Eq.(\ref{z})
is described by a single dimensionless function $ f(\theta) $
such that 
\beq
\label{bb7}
 \frac{ \partial^2}{ \partial \theta^2} \, \la 
\sigma \ra_{\theta} =   \frac{ \partial^2}{ \partial \theta^2} \,
\left( \la \sigma \ra_{0} \, f(\theta) \right) = \la \sigma \ra_{0}
\, f''(\theta) \; .
\eeq 
Any other form of introducing the $ \theta $ dependence can
be reduced to Eq.(\ref{bb7}). For example, the ansatz 
$ \la \sigma \ra_{\theta} = \la \sigma \ra_{0} 
f_{1}( \theta ) + \Lambda_{YM}^{4}
f_{2} (\theta) $ could be transformed to the form (\ref{bb7})
by the redefinition $  f_{1}( \theta ) +f_{2} (\theta) 
 \Lambda_{YM}^{4}/ \la \sigma \ra_{0} \rightarrow f(\theta) $. 
A function 
$ f(\theta) $ should satisfy the constraints $ f(0) = 1 \; , \; 
f'(0) = 0 $, which means that its small $ \theta $ expansion
reads
\beq
\label{bb8}
f(\theta) = 1 - 2 \xi^2 \, \theta^2 + \cdots \; ,
\eeq
where $ \xi^2 $ is some dimensionless number. Using this in 
Eq.(\ref{b7}), we obtain 
\beq
\label{bb9}
\lim_{ q \rightarrow 0} \; 
i \int dx \, e^{iqx} \lo T \left\{ \frac{\alpha_s}{8 \pi} 
G \tilde{G} (x)  \, 
\frac{\alpha_s}{8 \pi} G \tilde{G} (0) \right\}  \ro_{\theta = 0} =  
 \xi^2  \, \la -\frac{
 b \alpha_s}{ 8 \pi} G^2 \ra_{\theta = 0} \; .
\eeq
The assumption made in Eq.(\ref{5}) in the text was that 
Eq.(\ref{bb9}) is covariant in $ \theta $, i.e. remains of the 
same form (\ref{bb9}) not only for $ \theta = 0 $, but also for 
small values $ \theta \neq 0 $. This requirement fixes the function 
$ f(\theta) $ completely:
\beq
\label{bb10}
f''(\theta) = - 4 \xi^2 f(\theta) \; \Rightarrow \; 
f(\theta) = \cos (2 \xi \theta) \; , 
\eeq
which results in 
the $ \theta $ dependence of the 
vacuum energy and topological
density condensate identical to the one  
displayed in Eqs.(\ref{7}) and (\ref{8}). This assumption is 
self-consistent because the same Eqs.(\ref{7}),(\ref{8}) can 
be obtained without fixing the function $ f(\theta) $, but instead 
by postulating the covariant relation (\ref{5}) and resumming 
Taylor expansions in $ \theta $ for the condensates $ 
\la G^2 \ra_{\theta} \; , \; \la G \tilde{G} \ra_{\theta} $, as 
was done in \cite{6}.
One more relation needed for this purpose reads
\beq
\label{b8}
i \int dx \lo T \left\{ \frac{\alpha_s}{8 \pi} 
G^2 (x)  \, 
\frac{\alpha_s}{8 \pi} G \tilde{G} (0) \right\}  \ro =    
  \frac{4}{ b} \, \la \frac{
\alpha_s}{ 8 \pi} G \tilde{G} \ra \; ,
\eeq
which is a particular version of the original NSVZ 
theorem \cite{NSVZ}. Eq.(\ref{b8}) is valid for any small $ \theta $.

 Any other choice for the function
$ f(\theta) $, different from Eq.(\ref{bb10}), would presumably
not be self-consistent, though we did not find a general proof of 
this statement. One should stress that the form (\ref{bb10}), 
implying the reiteration of the parameter $ \xi^2 $ in multi-point
correlation functions of the operator $ Q $, is consistent with 
Veneziano construction \cite{Wit} for the ghost pole mechanism,
where it was found that
\beq
\label{bb11}
\frac{ \partial^{2n-1}}{ \partial \, \theta^{2n-1}} 
\la Q \ra \, |_{\theta= 0} \sim (-1)^{n-1} \,
N_c \, \left( \frac{ \lambda}{N_c}
\right)^{2n-1} \; \; , \; \; N_c \rightarrow \infty \; \; ,
\; \; \lambda = O(N_{c}^{0}) \; .
\eeq
At the same time, Eq.(\ref{bb10}) relates the ghost pole residue 
in Veneziano scheme  with conformal anomaly in the 
theory with $ \theta \neq 0 $, in terms of the 
parameters $ \xi \; , \la \sigma \ra_{0} $.     
 In our work \cite{6}, 
a particular choice, $ \xi = 4/(3 b ) $,
was advocated using  
a one-loop connection between the conformal 
and axial anomalies in the theory with an auxiliary heavy fermion.
Moreover, the covariance of Eq.(\ref{bb9}) in respect to $ \theta $
followed automatically within a procedure used in \cite{6}, 
where Eq.(\ref{bb9}) with $ \xi = 4/(3b) $ was obtained 
directly from the NSVZ relation (\ref{b3}) which is valid for 
any small $ \theta $.

\end{document}